\documentclass[conference]{IEEEtran}
\usepackage{amsmath}
\usepackage{hhline}
\usepackage[dvips]{graphicx} 
\usepackage{bmpsize} 
\usepackage{epsfig}
\usepackage{multirow}
\usepackage{setspace}
\usepackage{color}
\usepackage{url}
\usepackage{caption}
\usepackage{subcaption}
\usepackage{enumerate}
\usepackage{amsfonts}
\usepackage{amsthm}
\usepackage{amscd}
\usepackage{amssymb}
\usepackage[left=0.75in, right=0.75in, top=1in, bottom=0.75in]{geometry}

\begin{document}
	
	\title{Causal Models Applied to the Patterns of Human Migration due to Climate Change}
	
	\author{\IEEEauthorblockN{Kenneth Lai\textsuperscript{1,2} and Svetlana Yanushkevich\textsuperscript{1}}
		\IEEEauthorblockA{\textsuperscript{1}Biometric Technologies Laboratory, Department of Electrical and Software Engineering, University of Calgary, Canada\\ 
			\IEEEauthorblockA{\textsuperscript{2}Department of Clinical Neurosciences, Cumming School of Medicine, University of Calgary, Canada\\ 
				Email: \{kelai, syanshk\}@ucalgary.ca}}
	}	
	
	\markboth{IEEE SSCI,~2023}{ \MakeLowercase{\textit{et al.}}:
		.....}
	
	\maketitle
	
	\begin{abstract}
		The impacts of mass migration, such as crisis induced by climate change, extend beyond environmental concerns and can greatly affect social infrastructure and public services, such as education, healthcare, and security. These crises exacerbate certain elements like cultural barriers, and discrimination by amplifying the challenges faced by these affected communities. This paper proposes an innovative approach to address migration crises in the context of crisis management through a combination of modeling and imbalance assessment tools. By employing deep learning for forecasting and integrating causal reasoning via Bayesian networks, this methodology enables the evaluation of imbalances and risks in the socio-technological landscape, providing crucial insights for informed decision-making. Through this framework, critical systems can be analyzed to understand how fluctuations in migration levels may impact them, facilitating effective crisis governance strategies.
	\end{abstract}
	
	
	\section{Introduction}
	Mass migration, the movement of large groups of people from one region or country to another, is a pressing global challenge that has escalated in recent years due to factors such as conflict, climate change, economic disparities, and political instability. 
	
	There a numerous causes of mass migration with the few categories being (1) conflicts, (2) climate change, (3) economic disparities, and (4) demographic factors. Armed conflicts, civil wars, and political instability drive millions of people to flee their homes in search of safety. Rising sea levels, extreme weather events, and environmental degradation force communities to relocate in search of a more suitable environment. Economic inequality and lack of opportunities in certain regions compel individuals to seek better prospects elsewhere. Population growth and demographic shifts can lead to mass movements in search of resources and better living conditions.
	
	The impact of mass migration is in four main areas, particularly social/cultural, economic, environmental, and security.  Mass migration often leads to cultural exchanges, integration challenges, and potential tensions between host and migrant communities. Migrants can contribute positively to host economies through labor and skills, but they may also strain resources and welfare systems. Mass migration can lead to overexploitation of natural resources and increased pressure on ecosystems. Inadequately managed migration can raise security issues for both host and origin countries.
	
	Several challenges arise when dealing with mass migration: policy, infrastructure, services, and discrimination. Inconsistent and inadequate migration policies can hinder effective management and protection of migrants' rights.  Host regions may face difficulties in providing adequate housing, healthcare, and education to a sudden influx of migrants.  Deep-rooted prejudices can exacerbate tensions between local communities and migrants. Limited access to accurate migration data hampers evidence-based policymaking.
	
	\section{Problem Formulation and Related Work}
	There are many studies on the problem of migration \cite{andersson2016europe, lu2017systemic, estevens2018migration, lillywhite2022emergency}. One particular study by Andersson \cite{andersson2016europe} mentions that the mass migration crisis in Europe contains many challenges, in particular, how Europe's attempt to secure the border is a repetition of the same response resulting in a perpetual cycle.  The solution proposed is to break this vicious cycle by encouraging policymakers to directly address the underlying problems.
	
	Various solutions have been proposed to alleviate the problem of migration \cite{mcauliffe2021digitalization, cheesman2022self, corte2022blockchain, yanushkevich2019cognitive, bi2021integrated}. With the recent rise of artificial intelligence, \cite{mcauliffe2021digitalization} examines the increasing use of digital and AI technologies in the context of migration and evaluates the challenges and opportunities to migration due to these technologies.  In another case, \cite{cheesman2022self} examines the use of `self-soverign identity' (SSI), a user-controlled and decentralized form of digital identification, for border/migration management. SSI is closely tied to blockchain technologies where \cite{corte2022blockchain} reviews the use of blockchain technology in migrant systems to address problems of transparency and data consolidation.  A cognitive checkpoint has been proposed by \cite{yanushkevich2019cognitive} as an emerging technology that makes use of a biometric-enabled watchlist screening to fortify border control and security measures.
	
	This paper seeks to delve into the underlying impacts of mass migration, centering particularly on the potential impact of increased migration influx into Canada on the infrastructure and services of each province. The objective is to leverage machine learning models to predict future migration patterns, subsequently employing these projections to infer the implications for government infrastructure. We aspire to provide a comprehensive understanding of the interplay between migration dynamics and provincial capabilities, ultimately contributing to informed policy decisions and effective resource allocation.
	
	\section{Proposed Approach}
	In this paper, we propose an innovative approach that combines machine learning and machine reasoning techniques. Our approach leverages machine learning models to predict forthcoming migration trends by analyzing patterns within 7 years of historical migration data. Concurrently, we employ machine reasoning models based on causality networks, with a specific emphasis on Bayesian Networks. These reasoning models enable us to uncover the intricate distribution of current infrastructure and migration patterns across each province within Canada. Moreover, they facilitate an examination of how individual provinces provide support to migrants in various ways, encompassing economic aspects, sponsorships, and refugee assistance.
	
	By combining these machine learning and machine reasoning models, our methodology yields unique insights into the impacts of migration on each province. As migration numbers surge, our approach reveals the potential strain on existing infrastructure and services. In particular, we delve into scenarios where certain provinces may encounter challenges in accommodating heightened migration flows due to limitations in their infrastructure capacities.
	
	\section{Machine Learning Architecture for Forecasting}
	
	In this paper, we focus on 3 types of deep learning architecture to forecast migration numbers.  Transformer, Informer, and AutoFormer are all deep learning architectures and have seen significant uses in areas of natural language processing (NLP) and sequence-to-sequence tasks. In this paper, we used each of these architectures to forecast migration numbers when given monthly historic migration numbers from each province.  Each model is constructed similarly with 4 encoder layers, 4 decoder layers, and 32 dimensions for each layer.
	
	\subsection{Transformer}
	The first Transformer architecture was first introduced by \cite{vaswani2017attention}, with a focus on tackling natural language processing tasks.  In this paper, we repurpose this architecture for the task of forecasting time-series data, specifically monthly immigration to Canada.  The Transformer is built based on an encoder-decoder structure and is composed of a few key components that make it suitable for time-series-based tasks: self-attention, multi-head attention, positional encoding, and feedforward network.
	
	Self-attention is a mechanism that allows each data point in the input to be related to every other point.  This form of attention captures relationships between each point.
	
	Multi-head attention expands upon self-attention by performing the same self-attention task multiple times in parallel.  This allows the Transformer to connect to different parts of the input simultaneously.
	
	Positional encoding is a function of the Transformer that adds embeddings to the input to preserve information regarding sequence order.
	
	The feedforward network receives the information or the tokens from one attention layer and performs the processing to transform the output into a better representation as well as introduce non-linearity between each block in the Transformer.
	
	\subsection{Informer}
	The Informer architecture, proposed in  \cite{zhou2021informer} improves upon Transformer to address specific problems regarding time-series forecasting. This is done by modifying the Transformer to handle long sequences more efficiently via ProbSparse self-attention, self-attention distilling, and generative style decoder.
	
	ProbSparse self-attention is the proposed solution to address the problem of long sequences.  This mechanism shifts the processing from all time steps to focusing on the most important time steps, thus reducing the computational cost of processing the entire sequence.
	
	Self-attention distilling selects the most prominent and superior features and constructs a more concentrated self-attention feature map for the next attention layers.
	
	Generative style decoder attempts to address the problem of long predictions, particularly the impact on inference speed.  By predicting a long time-series sequence in one operation as opposed to dynamic decoding, this method of using generative inference reduces inference time for the decoder module.
	
	\subsection{AutoFormer}
	The AutoFormer architecture proposed in \cite{wu2021autoformer} also improves upon the Transformer architecture by introducing decomposition architecture with the auto-correlation mechanism.  These 2 techniques combined can be used to understand trends and seasonality of time-series data.
	
	Decomposition architecture is based on existing techniques of time-series decomposition, which separates a time-series into trend-cyclical and seasonal parts.  Authors \cite{wu2021autoformer} propose a series decomposition block to "extract the long-term stationary trend from predicted intermediate hidden variables progressively".
	
	Auto-Correlation is the mechanism used in Autoformer to find period-based dependencies by using a combination of autocorrelation and time delay aggregation.  
	
	\subsection{Metrics}
	In this paper, we characterize the performance of each of the forecast models in terms of MASE (Mean Absolute Scaled Error) \cite{hyndman2006another} and sMAPE (Symmetric Mean Absolute Percentage Error) \cite{makridakis1993accuracy}.  These two errors are common metrics used to measure the forecasting performance of models in time-series forecasting tasks.
	
	MASE (Eq. \ref{eq.mase}) is a scaled error metric that measures the accuracy of a model's predictions relative to a naive baseline forecast. It is useful for evaluating the forecast accuracy when dealing with non-seasonal time-series data.
	\begin{equation} \label{eq.mase}
		MASE=\frac{1/T \sum_{t=1}^{T} |y_t - y'_t|}{1/(T-1) \sum_{t=2}^{T} |y_t - y'_{t-1}|}
	\end{equation}
	where $T$ is the number of time steps in the test period, $y_t$ is the actual value at time step $t$ in the test set, and $y'_t$ is the predicted value at time step $t$ in the test set.
	
	The numerator computes the mean absolute error of the model's forecasts, while the denominator calculates the mean absolute error of a naive forecast, usually the one-step lag of the actual values. MASE compares the model's error to that of the naive forecast, and a value less than 1 indicates that the model performs better than the simple naive approach.
	
	sMAPE (Eq. \ref{eq.sMAPE}) is a percentage-based error metric that calculates the symmetric percentage difference between the actual and predicted values. It is useful when dealing with time-series data with different scales or magnitudes. 
	\begin{equation} \label{eq.sMAPE}
		sMAPE=1/T \sum_{t=1}^{T} \frac{2|y_t - y'_t|}{|y_t|+|y'_t|}
	\end{equation}
	where $T$ is the number of time steps in the test period, $y_t$ is the actual value at time step $t$ in the test set, and $y'_t$ is the predicted value at time step $t$ in the test set.
	
	sMAPE calculates the mean absolute percentage difference between the actual and predicted values and then symmetrizes the result to handle both underestimation and overestimation cases equally. The metric is bounded between 0\% and 200\%, with lower values indicating better forecast accuracy.
	
	Both MASE and sMAPE are widely used in evaluating the performance of forecasting models like Transformers as they provide meaningful insights into the accuracy and relative improvement over basic forecasts. These metrics help assess the effectiveness of different model configurations and fine-tune hyperparameters for better forecasting results.
	
	\section{Causal Network}
\begin{figure}[!h]
		\begin{center}
			\includegraphics[width=0.48\textwidth]{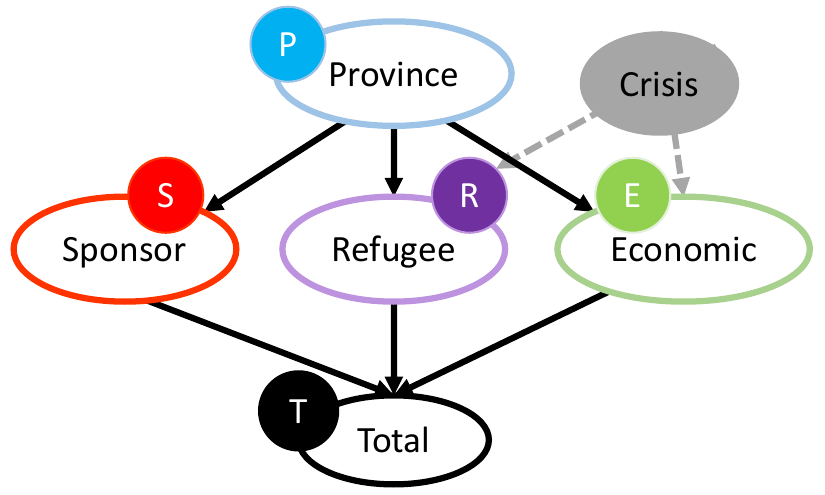}
			\caption{The proposed causal network consists of a hidden crisis node that influences the degree of migration, specifically impacting migration levels to `Refugee' and `Economic'.
			}\label{fig:causalNetwork}
		\end{center}
	\end{figure}

	Historically, significant disruptive events, such as the 2008 financial crisis, the 2015 EU migration crisis, the 2019 pandemic, and most recently the Russian-Ukraine conflict, have reshaped the socio-technological landscape, resulting in fluctuations in migration patterns. We provide a demonstrative experiment in order to understand and model the behavior of migration in response to these events.  We accomplish this by using a causal network (Fig. \ref{fig:causalNetwork}) comprising five nodes arranged in a diamond shape, with one parent, three child nodes, and one grandchild node. At the core of the network lies a unique hidden state representing the crisis event's impact, influencing changes in migration levels. While this hidden state is believed to influence refugee and economic nodes, its precise mechanism remains undisclosed due to the lack of comprehensive documentation. Through this causal network, we seek to shed light on the underlying dynamics that drive migration patterns during these disruptive events.
	
	This paper employs a probabilistic causal network to model immigration patterns in Canadian provinces based on immigration data from 2015 to 2023, sourced from the Government of Canada. The focus of this study is on the 10 provinces in Canada, while data from the territories is excluded due to relatively low migration levels.
	
	The network comprises four key nodes: `Province', `Sponsor', `Refugee', and `Economic', along with an aggregate node labeled `Total'. The `Province' node serves as the parent node, encompassing prior probabilities representing each province. We assume an equal distribution, assigning a 10\% probability to each province, British Columbia, Alberta, Saskatchewan, Manitoba, Ontario, Quebec, Newfoundland, New Brunswick, Prince Edward Island, and Nova Scotia.
	
	The remaining nodes, `Sponsor', `Refugee', `Economic', and `Total', quantify the average monthly migration (per 100 individuals) to each province. Specifically, `Sponsor' represents migration resulting from family sponsors, including children, spouses, and parents. `Refugee' denotes migration due to refugees or protected persons in Canada. `Economic' captures migration for economic reasons, such as worker programs, business ventures, or provincial nominee programs. Lastly, `Total' consolidates all types of migration to provide a comprehensive view of the overall immigration flow to each province.
	
	To represent the uncertainty associated with migration estimates, we employ probability distributions for each node, denoted by $\mu$ for average migration and $\sigma$ for one standard unit of standard deviation. For example, an $\mu=10$ in the `Economic' node indicates that, on average, 1000 (10*100) individuals migrate to the province monthly for economic reasons.
	
	By employing this probabilistic causal network, we aim to gain valuable insights into the dynamics of immigration patterns across Canadian provinces, facilitating informed decision-making and policy formulation in migration management.

	\section{Experimental Results}
	In this section, we present the experimental results of our study, which focuses on evaluating the performance of different deep learning architectures for time-series forecasting using the Mean Absolute Scaled Error (MASE) and Symmetric Mean Absolute Percentage Error (sMAPE) metrics. We explore the impact of varying context lengths on the forecasting accuracy of each Transformer model.
	
	In Table \ref{tab:Forecast}, we present the forecast results for each model with varying context lengths. Column (yr) represents the number of years of historical data used as context for each model. For example, a context length of 5 years utilizes 70 data points (12 months * 5 years) to forecast the migration for the next 12 months.
	
	Overall, the performance of the models is relatively similar across the various context lengths. For an 8-year context length, AutoFormer achieves the lowest MASE. Meanwhile, Informer exhibits the lowest sMAPE for both 6 and 8-year context lengths.
	
	\begin{table}[!ht]
		\caption{Forecast Performance for Autoformer, Transformer, and Informer at various context lengths (years).}
		\label{tab:Forecast}
		\begin{center}
			\small
			\begin{tabular}{c|cc|cc|cc}
				& \multicolumn{2}{c|}{Autoformer} & \multicolumn{2}{c|}{Transformer} & \multicolumn{2}{c}{Informer} \\
				
				(yr)  &	MASE	&	sMAPE	&	MASE	&	sMAPE	&	MASE	&	sMAPE	\\
				\hline
				
				1	&	2.1356	&	0.6121	&	2.3019	&	0.6091	&	2.1992	&	0.5927	\\
				2	&	2.0533	&	0.6083	&	2.0310	&	0.5763	&	1.9636	&	0.5666	\\
				3	&	1.9925	&	0.5832	&	2.0114	&	0.5705	&	1.9424	&	0.5734	\\
				4	&	2.0247	&	0.5819	&	2.0224	&	0.5824	&	1.9422	&	0.5677	\\
				5	&	2.0531	&	0.6055	&	2.1112	&	0.5943	&	1.9562	&	0.5736	\\
				6	&	1.9247	&	0.5797	&	2.0893	&	0.5858	&	1.9401	&	\textbf{0.5642}	\\
				7	&	2.0297	&	0.5810	&	2.1093	&	0.5926	&	2.0421	&	0.5854	\\
				8	&	\textbf{1.8999}	&	0.5823	&	2.0680	&	0.5869	&	1.9473	&	\textbf{0.5642}	\\
				9	&	1.9661	&	0.5797	&	2.1224	&	0.5977	&	1.9973	&	0.5930	\\
			\end{tabular}
		\end{center}
	\end{table}
	
	\section{Causal Analysis}
	Based on the proposed causal network (Fig. \ref{fig:causalNetwork}), we conducted an analysis using provincial immigration data to verify the impact of immigration through evidence and inference.
	
	\subsection{Case I: Ontario}
\begin{figure}[!htb]
		\begin{center}
			\includegraphics[clip, trim= 0 25 0 0, width=0.48\textwidth]{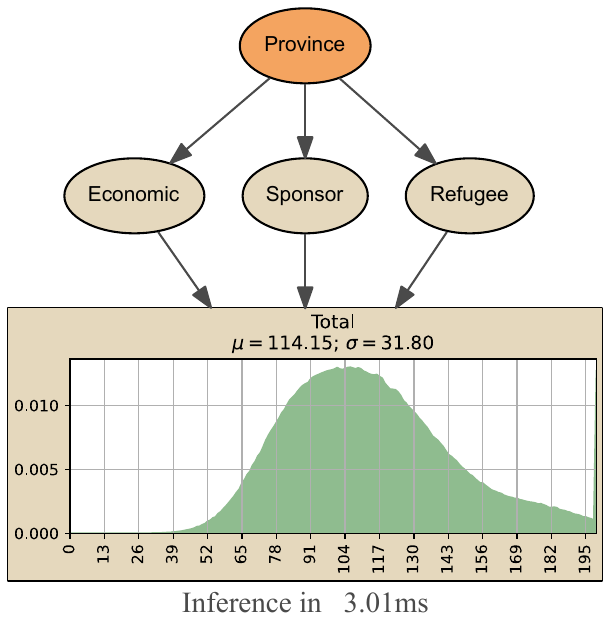}
			\caption{Causal network with provincial evidence, $Pr(\text{Total}|\text{Province}=\text{"Ontario"})$.  The conditional probability of `Total' is given to have approximately a $\mu=114.15$ and $\sigma=31.80$.
			}\label{fig:inferenceOntario}
		\end{center}
	\end{figure}

\begin{figure}[!htb]
		\begin{center}
			\includegraphics[clip, trim= 391 129 5 116, width=0.48\textwidth]{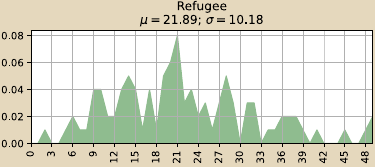}
			\caption{Probability distribution of `Refugee' with evidence, $Pr(\text{Refugee}|\text{Province}=\text{"Ontario"})$.
			}\label{fig:parts_province_inference}
		\end{center}
	\end{figure}

	We started by applying a diagnostic examination, setting the evidence to focus on the province of Ontario. In Fig. \ref{fig:inferenceOntario}, we illustrate the inference of the Bayesian network with the evidence Province=Ontario set. The resulting conditional probabilities represent a probability distribution. Specifically, the grandchild node Total' in Fig. \ref{fig:inferenceOntario} captures the distribution of immigration to Ontario, with an average level of 11,415 people per month and a standard deviation of 3,180. Additionally, the child node `Refugee' (Fig. \ref{fig:parts_province_inference}) represents the distribution of refugee immigration to Ontario, with an average level of 2,189 and a standard deviation of 1,018.
	
	\subsection{Case II: Refugee}
\begin{figure}[!h]
		\begin{center}
			\includegraphics[clip, trim= 169 100 5 77,width=0.48\textwidth]{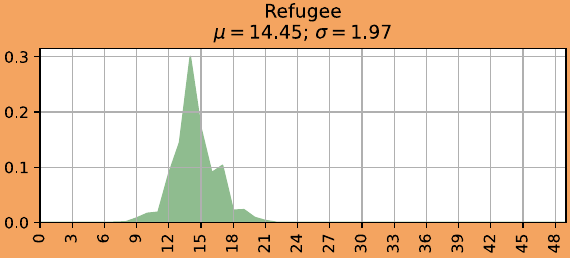}
			\caption{Refugee probability distribution, $\mathcal{N}(\mu=15,\sigma=2)$, provided as evidence to the network.
			}\label{fig:inferenceRefugee}
		\end{center}
	\end{figure}

	Next, we investigate how each node changes when the refugee level is modified. Assuming a refugee distribution of 1,445 with a standard deviation of 197 (approximately $\mathcal{N}(\mu=15, \sigma=2)$), we use this as evidence for refugee immigration (Fig. \ref{fig:inferenceRefugee}) and infer its causal impact on the probabilities of `Province' and `Total.'
	
\begin{figure}[!htb]
		\begin{center}
			\includegraphics[clip, trim= 246 238 248 3, width=0.24\textwidth]{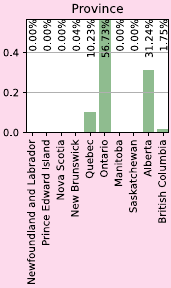}\\
			\caption{Probability distribution of ``Province'' with ``Refugee'' evidence, $Pr(\text{Province}|\text{Refugee}\approx \mathcal{N}(\mu=15,\sigma=2))$.
			}\label{fig:province_refugee_inference}
		\end{center}
	\end{figure}

\begin{figure}[!htb]
		\begin{center}
			\includegraphics[clip, trim= 195 22 197 274, width=0.48\textwidth]{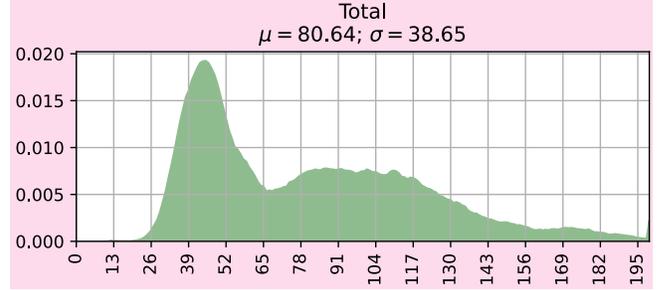}\\
			\caption{Probability distribution of ``Total'' with ``Refugee'' evidence, $Pr(\text{Total}|\text{Refugee}\approx \mathcal{N}(\mu=15,\sigma=2))$.
			}\label{fig:total_refugee_inference}
		\end{center}
	\end{figure}

	In Fig. \ref{fig:province_refugee_inference} and Fig. \ref{fig:total_refugee_inference}, we illustrate the causal impact as a result of the refugee evidence. It shows that this level of refugee immigration will directly impact five provinces: New Brunswick, Quebec, Ontario, Alberta, and British Columbia. The province most influenced is Ontario with 56.73\% impact, followed by Alberta with 31.24\%, and Quebec with 10.23\%.
	
	The immigration distribution for the `Total' node indicates that on average, 8,064 people will be joining the province. When compared to Ontario's monthly intake of 11,415, this indicates a lower level of immigration, resulting in less stress on services and facilities in Ontario. However, when compared to the monthly intake of other provinces (e.g., Alberta: 3,568, Quebec: 4,099, British Columbia: 3,949, New Brunswick: 434), the average immigration to `Total' is nearly double, indicating a potential strain on services and facilities in the mentioned provinces. In such cases, additional resources may be required to alleviate the stress and support the incoming migrants.
	
	These results offer valuable insights into the causal relationships between immigration and provinces, allowing for more informed decision-making in immigration management and resource allocation during migration crises.
	
	\begin{figure*}[!htb]
		\begin{center}
			\includegraphics[clip, trim= 0 15 0 0, width=\textwidth]{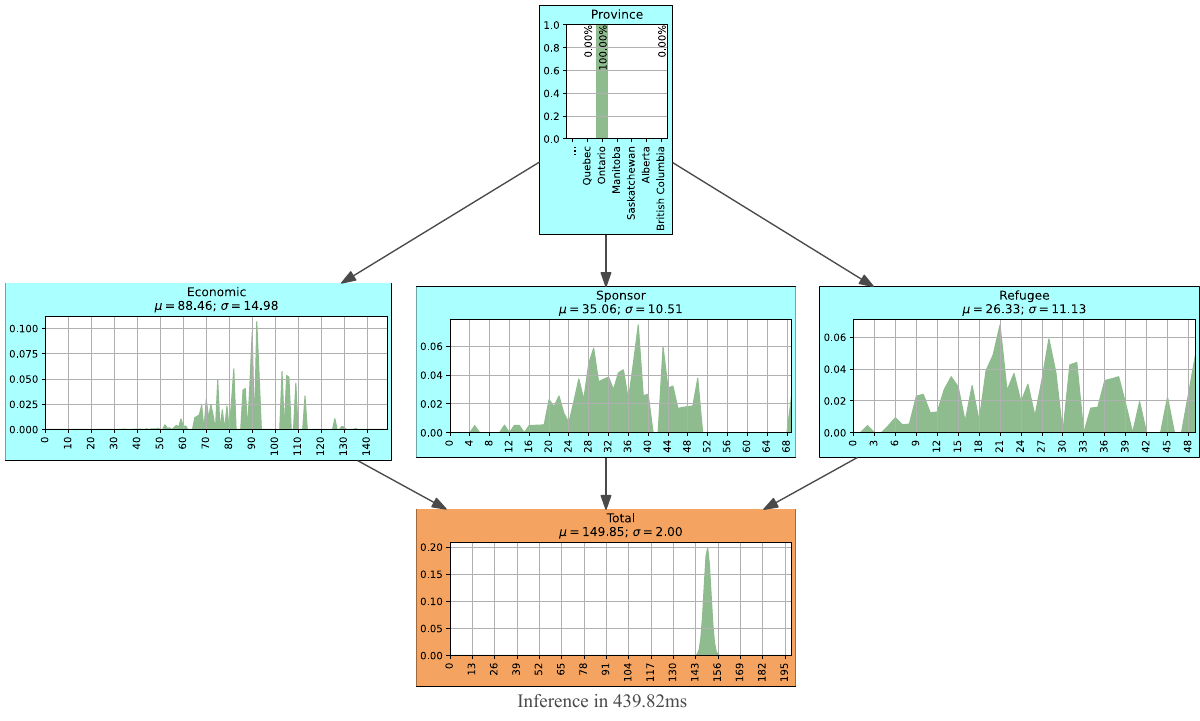}
			\caption{Probability distributions with forecasted evidence of 15,000 people per month, $Pr(\text{Total}=\mathcal{N}(\mu=150,\sigma=2))$.
			}\label{fig:total_inference}
		\end{center}
	\end{figure*}
	
	\subsection{Case III: Total}
	Following the forecasts made by deep learning models, Fig. \ref{fig:total_inference} presents an assumption that the total migration to any province amounts to 15,000 individuals monthly. Under this high level of immigration, the results indicate that only Ontario, without any additional resource allocation, would be capable of accommodating the influx. The inference also reveals that among the 15,000 migrants, 8,846 would be due to economic reasons, 3,506 due to family sponsors, and 2,633 would be refugees.
	
	These findings highlight a concerning trend: as immigration levels to Canada increase, the existing facilities and services become insufficient to provide adequate support for all individuals. Proactive measures are imperative to enhance the level of support available to ensure that each province can share the load effectively.
	
	In summary, the forecast and inference results signal the necessity for strategic planning and resource allocation to address the challenges posed by rising immigration levels. It is crucial for policymakers to take proactive actions to bolster support systems, ensuring a sustainable and equitable distribution of resources across provinces during migration crises.
	
	\section{Discussion and Conclusion}
	
	In this study, we proposed a comprehensive approach to forecast migration patterns and analyze their causal relationships using deep learning models and Bayesian networks. We focused on three state-of-the-art Transformers: the original Transformer, Informer, and AutoFormer, to forecast migration trends in response to historical events during various context lengths. Additionally, we employed a Bayesian network to investigate the causal impact of immigration on specific provinces, shedding light on the factors influencing migration patterns.
	
	The experimental results demonstrated the effectiveness of the different Transformers in forecasting migration trends. While all models exhibited competitive performance across various context lengths, Informer marginally outperformed the others. Furthermore, our causal analysis using the Bayesian network provided valuable insights into the relationships between historical events and migration dynamics. The analysis revealed the significant impact of economic factors, refugee crises, and government policies on migration patterns. These findings are crucial for policymakers and stakeholders in developing targeted interventions and crisis governance strategies.
	
	Our research contributes to the field of migration forecasting and crisis management by offering interpretable models and causal analysis. The integration of deep learning models and Bayesian networks provides a powerful framework for understanding migration patterns and their drivers, facilitating more informed decision-making in migration governance.
	
	While our study presents promising results, there are still opportunities for future research. Exploring the incorporation of additional features, such as socio-economic indicators and climate data, could further enhance the forecasting accuracy of the models. Moreover, expanding the scope of the Bayesian network to include more factors and provinces would deepen our understanding of migration dynamics on a larger scale.
	
	\subsection*{Acknowledgment}
	
	\begin{small}
		This work was supported in part by the Social Sciences and Humanities Research Council of Canada (SSHRC) through the Grant ``Emergency Management Cycle-Centric R\&D: From National Prototyping to Global Implementation'' under Grant NFRF-2021-00277; in part by the University of Calgary under the Eyes High Postdoctoral Match-Funding Program.
	\end{small}

	{\small
		\bibliographystyle{IEEEtran}
		\bibliography{bib}
	}
	
\end{document}